\documentclass{mem}
\usepackage{natbib}\usepackage{txfonts}\usepackage{balance}
\usepackage{graphicx}
\usepackage[a4paper]{hyperref}
\begin{document}

\title{Anisotropic AGN Outflows Filling The Cosmological Volume}

\author{Paramita Barai\inst{1}, 
Jo\"el Germain\inst{1} 
\and Hugo Martel\inst{1}}

\institute{
D\'epartement de physique, de g\'enie physique et d'optique,
Universit\'e Laval, Qu\'ebec, QC, Canada
\email{paramita.barai.1@ulaval.ca}
}
%\altaffiltext{2}{Centre de Recherche en Astrophysique du Qu\'ebec}

\authorrunning{Barai et al.}

\titlerunning{Cosmological Impact of AGN Outflows}

\abstract{
%We seek to compute the fraction of the volume of the Universe filled by 
% expanding cocoons of as well as the magnetic field infused by them, 
We simulate anisotropic outflows of AGN, 
and investigate the large-scale impact 
of the cosmological population of AGN outflows over the Hubble time 
by performing N-body $\Lambda$CDM simulations. %of $\Lambda$CDM Universe. 
%We compute the cosmological volume filled by anisotropic AGN outflows, 
Using the observed quasar luminosity function to get the redshift and luminosity distribution, 
and analytical models for the outflow expansion, 
AGNs are allowed to evolve in a cosmological volume. 
By the present epoch, $13 - 25\%$ of the total volume %of the Universe 
is found to be pervaded by AGN outflows, with $10^{-9}$ G magnetic field. 
} 
\maketitle{}

\section{Introduction} 
\label{sec-intro} 

Outflows from AGN are observed in a wide variety of forms: 
radio galaxies, broad absorption line quasars, 
Seyfert galaxies exhibiting intrinsic absorption in the UV, broad emission lines, 
warm absorbers and absorption lines in X-rays 
\citep[e.g.,][]{crenshaw03, everett07}. 
There have been studies on the cosmological impact of quasar outflows in large scales 
(\citealt{FL01}, hereafter FL01; \citealt{so04}, hereafter SO04; \citealt{lg05}, hereafter LG05). 
\citet{barai08} investigated the cosmological influence of radio galaxies over the Hubble time. 
All these studies considered spherically expanding outflows. 

On cosmological scales an outflow is expected to move away from 
the high density regions of large-scale structures, with the outflowing matter 
getting channelled into low-density regions of the Universe \citep{martel01}. 
%which causes an outflow to %attaining an anisotropic shape.
We implement such anisotropic AGN outflows within a cosmological volume. 
The simulation methodology is given in \S\ref{sec-numerical}, 
and the results are discussed in \S\ref{sec-results}. 

\section{The Numerical Setup} 
\label{sec-numerical} 

\subsection{N-body Simulation and Distribution}
\label{sec-num-Nbody} 

We simulate the growth of large-scale structures in a cubic cosmological
volume of comoving size $L_{\rm box}$=$128 h^{-1}$ Mpc. 
We use the Particle-Mesh (PM) algorithm, %\citep{he88}. 
with $256^3$ equal-mass particles, on a $512^3$ grid. 
A particle has a mass of $1.32\times10^{10} M_\odot$, 
and the grid spacing is $\Delta = 0.25 h^{-1}$ Mpc. 
We consider a concordance $\Lambda$CDM model 
with the cosmological parameters: 
$\Omega_M=0.268$, $\Omega_{\Lambda}=0.732$, $H_0=70.4\,\rm km\,s^{-1}Mpc^{-1}$,
$\Omega_b=0.0441$, $n_s=0.947$, and $T_{\rm CMB}=2.725$. 
%consistent with the results of {\sl WMAP3}. %\citep{spergel07}. 

%\subsection{Luminosity and Spatial Distribution} 
%The AGNs are distributed in the cosmological volume as given below. 

The redshift-dependent luminosity distribution of AGN is obtained from the %work on 
bolometric quasar luminosity function (QLF) \citep{hopkins07}, 
%The QLF is expressed as a standard double power law,
\begin{equation}
\phi(L,z) \equiv \frac{d\Phi}{d \log L} 
= \frac{\phi_{\star}} {(L/L_{\star})^{\gamma_1} + (L/L_{\star})^{\gamma_2}}, 
\end{equation} 
which gives the number of quasars per unit comoving volume, 
per unit $\log_{10}$ of luminosity. 
%We adopt the values of parameters $\phi_{\star}$, $L_{\star}$, 
%$\gamma_1$ and $\gamma_2$ from Table 2 of \citet{hopkins07}. 
A fraction $f_{\rm outflow}= 0.2$ of AGN are considered to host outflows \citep{ganguly08}.
The number of %AGN with 
outflows within the simulation box of comoving volume $V_{\rm box} = L_{\rm box}^3$, 
at epoch $z$, %in the $L$ interval 
between $[L, L+dL]$ is, 
\begin{equation}
N(L,z) = f_{\rm outflow} \phi(L,z) d[\log_{10} L] V_{\rm box}.
\label{eq-N-Lz}
\end{equation} 
The AGN activity lifetime is taken as $T_{\rm AGN} =10^8$ yr; 
the maximum and minimum AGN luminosities as 
$10^{8} L_{\odot}$ and $10^{14} L_{\odot}$. %\citep{crenshaw03}. 

Using the QLF, we obtain the entire cosmological population of AGN 
in the simulation volume starting from $z = 6$, namely the birth redshift ($z_{\rm bir}$), 
switch-off redshift ($z_{\rm off}$) and bolometric luminosity ($L_{\rm bol}$) of each source. 
A total of 929805 sources were produced. 

At each timestep, 
we filter the density distribution on the $512^3$ grid (from the PM code) 
using a gaussian filter containing a mass $10^{10} M_{\odot}$, 
assumed as the minimum mass of a halo hosting an AGN. 
We identify the density peaks, or 
the grid cells where the filtered density exceeds the values 
at the 26 neighboring grids. %are the density peaks, 
%where we spatially locate the new AGNs born during that epoch 
%(whose $z_{\rm bir}$ values fall within the timestep interval). 
We consider the peaks that have a filtered density $> 5 \times$ the mean density of the box, 
and each new AGN born during that epoch 
(whose $z_{\rm bir}$ values fall within the timestep interval) 
is located at the center of one such peak cell, selected randomly. 
%We also check for overlap and do not locate a new source in a cell 
%which already contains a AGN born at an earlier epoch. 
After their initial distribution, the AGNs are allowed to evolve according to 
the prescription in \S\ref{sec-num-outflow}. 
%Hence we compute the volume of the cosmological box filled by 
%the expanding AGN outflows. 

\subsection{Outflow Model} 
\label{sec-num-outflow} 

Despite the observational differences between various outflows, 
%the important point relevant for the present study is 
we stress that the AGNs hosting outflows 
constitute a random subset of the whole AGN population, %(SO04), 
and we simply assume the same outflow model 
for all AGNs (also FL01, SO04, LG05). 
We allow each outflow to evolve through an active-AGN life 
(\S\ref{sec-num-active}), when $z_{\rm bir} > z > z_{\rm off}$. 
After the central engine has stopped activity (when $z < z_{\rm off}$), 
it enters the late-expansion phase (\S\ref{sec-num-aniso}). 

The baryonic ambient gas density, $\rho_g(z, {\bf r})$, 
%forming the ambient medium for the outflows, 
is considered to follow the dark matter density, $\rho_{M}(z, {\bf r})$, 
in the N-body simulation: 
$\rho_g = (\Omega_b / \Omega_M) \rho_{M}$. 
The external gas pressure is 
$p_g(z, {\bf r}) = \rho_g(z, {\bf r}) k T_g/ \mu$. 
The external temperature is fixed at $T_g = 10^4$ K assuming 
a photoheated ambient medium, and $\mu = 0.611$ amu is the mean molecular mass. 

\subsubsection{The Active Life} 
\label{sec-num-active} 

The AGN activity period is short compared to the Hubble time. 
So we neglect energy losses and Hubble expansion 
of the cosmological volume when the quasar is active. 
We approximate the shape of the outflow %expanding during the active AGN life 
as spherical. %with radius $R$. 

An active outflow is inflated by twin collimated relativistic jets %expanding from the central AGN 
\citep{begelman84}, each of length $R$. 
%According to the standard scenario \citep[e.g.,][]{begelman89}, 
%the cocoon pressure $p_c$ is much larger than the ambient gas pressure. 
We consider that the kinetic luminosity carried by each jet 
is a constant fraction of the bolometric luminosity: 
$L_K = \epsilon_K L_{\rm bol} / 2$, with %the kinetic fraction as 
$\epsilon_K = 0.1$ (FL01; LG05; \citealt{shankar08}). % chartas07 
%%SO04 uses $\epsilon_K = 0.05$. 
The jet advance speed %$v_s$ 
%of the shock at the jet head can be 
is obtained by balancing the jet momentum flux with the ram pressure of the ambient medium: 
$L_K / (A_s c) = \rho_g (dR / dt)^2$. 
Here, $A_s$ is the area of the shocked ``working'' surface at the jet head. 
We use $A_s  = 2 \pi R^2 \theta_s^2$, 
assuming that the shock front has a constant half-opening angle of 
$\theta_s = 5^{\circ}$ relative to the central AGN (FL01). 

All the kinetic energy transported along the jets during an AGN's age 
$t_{\rm age} = t(z) - t(z_{\rm bir})$ is transferred to the outflow. 
The outflow energy %within the outflow 
is $E_0 = 2 L_K t_{\rm age}$, and 
its pressure follows $p_0 V_0 = (\Gamma_0-1) E_0$. 
The adiabatic index of the relativistic outflow plasma is $\Gamma_0  = 4/3$. 
%We approximate the shape of the outflow expanding during 
%the active AGN life as spherical with radius $R$. 
%The equation of motion is $dR / dt = v_s$, 
%which we solve using a 4th-order Runge-Kutta integration method. 
%solving which we get 
The outflow volume during this active spherical expansion is $V_0 (z) = 4 \pi R^3 / 3$. 
% V_{\rm outflow}

%\subsubsection{The Dormant Phase: Anisotropic Expansion} 
\subsubsection{The Late Anisotropic Expansion} 
\label{sec-num-aniso} 

When AGN activity ends, 
the left-over high-pressure outflow expands into the large scales 
of the IGM with an anisotropic geometry. 
%The pressure inside the outflow causes it to continue expanding as long as it is overpressured. 
%An anisotropic outflow is 
It is represented as a ``bipolar spherical cone'' 
with radius $R$ and opening angle $\alpha$ \citep{pmg07}, 
and it follows the direction of least resistance (DLR). %as it travels away from the source. 
We perform a second-order Taylor expansion of the density around each peak, 
whose coefficients are determined by performing a least-square fit 
to all the grid points within a distance $2 \Delta$ from the peak. 
We then rotate the %Cartesian 
coordinate axes %to $(x', y', z')$, 
such that the cross-terms vanish to give: 
$\rho(x', y', z')=\rho_{\rm peak} - A{x'}^2 - B{y'}^2 - C{z'}^2$. 
The largest of the coefficients $A$, $B$, $C$ gives the DLR. 
%direction along which the density drops the fastest as we move away from the peak. 
%We take it as the DLR.

%This overpressured outflow expansion occurs 
The Sedov-Taylor adiabatic blast wave model is used to obtain 
the radius of the overpressured outflow (\citealt{castor75}; SO04; LG05): 
$R (z) = \xi_0 \left( E_{\rm tot} t_{\rm age}^2 / \overline{\rho_g}(z) \right) ^ {1/5}$. 
We obtain $\overline{\rho_g} (z)$ by averaging the gas density of the grid cells 
%in the simulation box 
occurring within the outflow volume. 
For a strong explosion in the $\Gamma_g = 5/3$ ambient gas, $\xi_0 = 1.12$. 
Here the total kinetic energy injected into the outflow by the AGN 
%throughout the active lifetime 
is $E_{\rm tot} =  2 L_K T_{\rm AGN}$. 
%The outflow is considered to undergo a 
Adiabatic expansion losses are considered, and 
the outflow pressure evolves as $p_0 R^{3 \Gamma_g} =$ constant, 
with the constant derived from the %pressure and size it had at the end of the active phase. 
values at the end of the active phase. 

The outflow follows an anisotropic expansion as long as its pressure exceeds 
the external pressure of the IGM, i.e., $p_0(z) > p_g(z)$. 
During this late biconical expansion, $V_0 (z) = 4 \pi R^3 (1 - \cos (\alpha/2)) / 3$. 

When $p_0(z) \leq p_g(z)$, or, the outflow has reached pressure equilibrium 
with the IGM, it has no further %intrinsic 
expansion. After this, the outflow simply evolves passively with the 
Hubble flow of the cosmological volume. 
Thus an outflow in pressure equilibrium attains a final volume of 
$V_0 = 4 \pi R_f^3 (1 - \cos (\alpha/2)) / 3$, where $R_f$ is the final comoving radius of the outflow. 

\section{Results and Discussion} 
\label{sec-results}

% FIGURE 1 

\begin{figure}[]
\resizebox{\hsize}{!}{\includegraphics[clip=true]{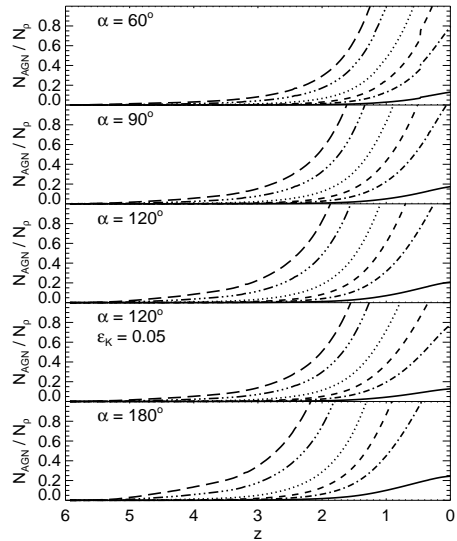}}
\caption{ 
\footnotesize
Volume of the simulation box filled by AGN outflows ($N_{\rm AGN}$) 
as a fraction of the total volume {\it (solid)}, 
and as a fraction of volumes of various overdensities: 
$N (\rho > \overline{\rho})$ {\it (dash dot)}, $N (\rho > 2 \overline{\rho})$ {\it (dashed)}, 
$N (\rho > 3 \overline{\rho})$ {\it (dotted)}, $N (\rho > 5 \overline{\rho})$ {\it (dash dot dot dot)}, 
$N (\rho > 7 \overline{\rho})$ {\it (long dashes)}. 
%The panels from top to bottom are for 
%opening angles of  $\alpha = 60, 90, 120, 120, 180^{\rm o}$. 
%All uses $\epsilon_K = 0.1$, except the 4th panel where $\epsilon_K = 0.05$. 
}
\label{fig1} 
\end{figure} 

At each timestep the total volume occupied by the AGN outflows 
is computed by counting the contributions of all the sources born by then, 
both the active ones and those in the anisotropic phase. 
%This gives the redshift evolution of the total AGN volume in the simulation box. 
We performed 4 simulations with opening angles of 
$\alpha = 60^{\rm o}, 90^{\rm o}, 120^{\rm o}, 180^{\rm o}$, 
all with $\epsilon_K = 0.1$, and one with 
$\alpha = 120^{\rm o}$ and $\epsilon_K = 0.05$, 
whose results are shown in the figures. 

%In order to prevent overcounting of the volume due to overlap of outflows, 
We count the grid cells in the simulation box which occur inside 
the volume of one or more AGN outflows. 
The total number of these filled cells, $N_{\rm AGN}$, 
gives the total volume of the box occupied by outflows. 
We express the total volume filled as a fraction of volumes of 
various overdensities in the box, $N_{\rho} = N (\rho > {\cal C} \overline{\rho})$, 
where $\overline{\rho} = (1+z)^3 \Omega_M 3H_0^2 / (8 \pi G)$ 
is the mean matter density of a spatially flat Universe (the box) at an epoch $z$. 
So $N_{\rho}$ gives the number of cells which are at a density ${\cal C}$ times the mean density. 
We find $N_{\rho}$ for ${\cal C} = 0, 1, 2, 3, 5, 7$; ${\cal C} = 0$ gives the total volume of the box. 
%since then $N_{\rho} = N (\rho > 0) = 512^3$ is the total number of cells in the box. 

Figure~\ref{fig1} shows the redshift evolution of the volume filling factors for our 5 simulation runs. 
%when $\epsilon_K = 0.1$, 
With $10\%$ kinetic efficiency, 
$0.13$ of the entire Universe is filled at present %$z = 0$ 
by AGN outflows with an opening angle of $60^{\rm o}$; 
the fraction increases to $0.17$ with $90^{\rm o}$, $0.21$ with $120^{\rm o}$, and 
$0.25$ with $180^{\rm o}$. 
%With $\epsilon_K = 0.05$, it is $0.13$ with $120^{\rm o}$. 
%By halving $\epsilon_K$, the filled fraction becomes ... 
A $5\%$ kinetic efficiency and $\alpha = 120^{\rm o}$ fills $0.13$ of the volume. 
In all our runs, the outflows fill up all of the regions with $\rho > 2 \overline{\rho}$ %or higher 
by $z=0.3$. %or before. 
With $\epsilon_K = 0.1$ and $\alpha = 90^{\rm o}$ or higher, 
the outflows permeate all the overdense regions ($\rho > \overline{\rho}$) by $z = 0.1$. 

% FIGURE 2 

\begin{figure}[]
\resizebox{\hsize}{!}{\includegraphics[clip=true]{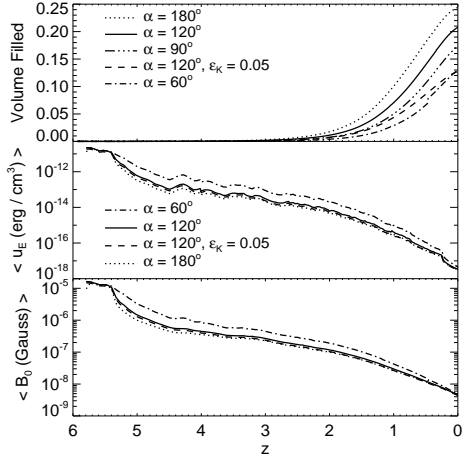}}
\caption{ 
\footnotesize 
Volume fraction filled by AGN outflows ({\it top}), 
the volume weighted average of the total energy density inside outflow volumes 
$\langle u_E \rangle$ ({\it middle}), 
and the equipartition magnetic field within the filled volumes 
$\langle B_0 \rangle$ ({\it bottom}). 
%Linestyles indicate different parameters. 
}
\label{fig2} 
\end{figure} 

It is the overdense regions of the Universe 
which gravitationally collapse to form stars and galaxies. 
So evidently the AGN outflows have a profound cosmological impact 
on the protogalactic regions. 
We note that the volumes obtained by LG05 (100\% filling by $z\sim1$) 
are higher than ours. 

We perform preliminary estimates of the energy density and 
magnetic field in the volumes of the Universe filled by the AGN outflows. 
The energy density inside the outflow behaves similar to the outflow pressure 
evolving adiabatically (\S\ref{sec-num-aniso}), $u_E = 3 p_0$. 
Assuming equipartition of energy between magnetic field of strength $B_0$ 
and relativistic particles inside the outflow, 
the magnetic energy density is $u_B = u_E / 2 = B_0^2 / (8 \pi)$. 
%The mean thermal energy density of the ambient medium inside the outflow volume is 
%$\overline{u_{T,g}} = 3 \overline{\rho_g} k T_g / (2 \mu)$. 
We define the volume weighted average of a physical quantity ${\cal A}$ as 
$\langle {\cal A} \rangle (z) \equiv \sum ({\cal A} V_0) / \sum V_0$, 
where the summation is over all outflows existing in the simulation box at that epoch. 

Figure~\ref{fig2} shows the redshift evolution of 
the total volume filling fraction, $\langle u_E \rangle$ and $\langle B_0 \rangle$. 
The energy density and magnetic field decrease with redshift 
as larger volumes are filled. 
At $z = 0$, a magnetic field of $\sim 10^{-9}$ G permeates the filled overdense volumes, 
consistent with the results of \citet{ryu08}. 
At a given redshift, 
the energy density and magnetic field are larger for smaller opening angles 
of the anisotropic outflows. 

%\section{Conclusion} \label{sec-conclusion} 

We conclude that, using our N-body simulations, 
the cosmological population of AGN outflows pervade $13-25\%$ 
of the volume of the Universe by the present. 
%and occupy $100\%$ of the overdense regions by $z \sim 0.3$. 
A magnetic field of $\sim 10^{-9}$ G is infused in the filled volumes at $z = 0$.

\begin{acknowledgements} 
All calculations were performed at the Laboratoire 
d'astrophysique num\'erique, Universit\'e Laval. 
We thank the Canada Research Chair program and NSERC for support. 
\end{acknowledgements}

%%%%%%%%%%%%%%%%%%%%%%%%%%%%%%%%%%%%%%%%%
%                   REFERENCES

\bibliographystyle{aa}

\end{document}